\documentclass{iopart}

\usepackage[T1]{fontenc}

\usepackage{graphicx}
\usepackage{epstopdf}
\usepackage{amssymb}
\usepackage{tikz}
\usetikzlibrary{decorations.pathmorphing}
\usetikzlibrary{arrows,shapes}
\usepackage{subfigure}
\usepackage[urlcolor=blue,hyperindex,colorlinks,bookmarks=true,linkcolor=black,citecolor=black]{hyperref}
\usepackage[normalem]{ulem}
\usepackage{color}
\usepackage{rotating}
\usepackage{enumerate}

\newcommand{\ket}[1]{\left|#1\right\rangle}

\begin{document}

\title{Quantum Simulation of a Discrete-Time Quantum Stochastic Walk}

\author{Peter K. Schuhmacher$^1$, Luke C. G. Govia$^{1,2}$, Bruno G. Taketani$^{1,3}$ and Frank K. Wilhelm$^1$}
\address{$^1$ Theoretical Physics, Saarland University, Campus, 66123 Saarbr\"{u}cken, Germany}
\address{$^2$ Raytheon BBN Technologies, 10 Moulton Street, Cambridge, MA, 02138, USA}
\address{$^3$ Departamento de F\'isica, Universidade Federal de Santa Catarina, 88040-900, Florian\'opolis, Brazil.}

\begin{abstract}
Quantum walks have been shown to have a wide range of applications, from artificial intelligence, to photosynthesis, and quantum transport. Quantum stochastic walks (QSWs) generalize this concept to additional non-unitary evolution. In this paper, we propose a trajectory-based quantum simulation protocol to effectively implement a family of discrete-time QSWs in a quantum device. After deriving the protocol for a 2-vertex graph with a single edge, we show how our protocol generalizes to a graph with arbitrary topology and connectivity. The straight-forward generalization leads to simple scaling of the protocol to complex graphs. Finally, we show how to simulate a restricted class of continuous-time QSWs by a discrete-time QSW, and how this is amenable to our simulation protocol for discrete-time QSWs.
\end{abstract}

\maketitle

\section{Introduction}
\label{sec:intro}

The quantum mechanical analogue to the ubiquitous classical random walk on a graph is the so-called quantum walk~\cite{Aharonov1993}. Quantum walks can be either continuous-time~\cite{Farhi1998}, or discrete-time~\cite{Aharonov2001,ambainis2001one}, and as both versions have been shown to be universal for quantum computation~\cite{Childs2003,Lovett2010} they offer a powerful paradigm for both studying, and harnessing quantum mechanics for computational applications. Examples of this include machine learning~\cite{Schuld2014c,Rebentrost2014}, search algorithms~\cite{Shenvi2003} and photosynthetic excitation transfer~\cite{Mohseni2008,Walschaers:2013aa}.

Quantum walks are completely coherent and hence, the walk is naturally reversible and undirected, as it follows Hamiltonian dynamics. In directed walks, time-reversal symmetry is broken as vertices can be connected by one-way edges. This condition implies that these walks are described by non-Hermitian dynamics and thus cannot be directly implemented in a quantum computer. Quantum stochastic walks (QSWs) are a simple way to represent such evolutions as they combine both quantum unitary walks and non-unitary, stochastic evolution~\cite{Whitfield2010}. In the continuous-time case it was recently shown that the reservoir engineering required to implement such walks is a problem as hard as the problem the QSW is designed to solve~\cite{Taketani2016}, though in some cases a quantum simulation approach can be taken~\cite{Govia2017}. For discrete-time QSWs the scenario is different as one can take advantage of measurement-based feed-forward schemes to implement the directionality~\cite{Schuld2014c}.

In this work we propose an algorithm to simulate discrete-time QSWs. The central concept behind our protocol is that if one performs randomly chosen unitary dynamics from a carefully designed set, this can implement a specific non-unitary evolution in the ensemble average~\cite{CrispinGardiner2004}. Our protocol is based on ancilla systems and a feed-forward scheme to implement the required evolution. The simplicity of the implementation of a single edge lends to straight-forward scaling to more complex graphs, and is a key feature of the protocol.

This paper is organized as follows. In Section~\ref{sec:Kraus} we present the ensemble average formalism using the Kraus decomposition. Section~\ref{sec:Protocol} describes the algorithm, starting from a 2-vertex directed graph and generalizing it to complex graphs. In section \ref{sec:ApproximateTimeEvolution}, we show how to simulate a restricted class of continuous-time QSWs via a discrete-time QSW. Finally, Section~\ref{sec:conc} presents our concluding remarks.

\section{Quantum Simulation of a Kraus Map}
\label{sec:Kraus}
The algorithm proposed here will be formulated on the Kraus decomposition of the desired QSW. Any completely-positive and trace preserving quantum operation~\cite{Nielsen2000}, can be written in Kraus operator form as
\begin{eqnarray}
        \mathcal{B}[\rho] = \sum_j \hat{K}_j\rho\hat{K}_j^\dagger,
\end{eqnarray}
where $\{\hat{K}_j\}$ are the Kraus operators, which must satisfy $\sum_j\hat{K}_j^\dagger\hat{K}_j = \hat{\mathbb{I}}$ to preserve the trace of the quantum state. This condition implies that
\begin{eqnarray}
 {\rm Tr}\left[\sum_j\hat{K}_j^\dagger\hat{K}_j\rho\right] = 1~~\Rightarrow&\sum_j{\rm Tr}\left[\hat{K}_j^\dagger\hat{K}_j\rho\right] = \sum_j\tilde{P}_j = 1,
\end{eqnarray}
where we have defined the probabilities $\tilde{P}_j = {\rm Tr}[\hat{K}_j^\dagger\hat{K}_j\rho]$, which are guaranteed to be non-negative as $\hat{K}_j^\dagger\hat{K}_j$ is Hermitian. We can then rewrite our original quantum operation as
\begin{eqnarray}
\mathcal{B}[\rho] = \sum_j \tilde{P}_j\tilde{K}_j\rho\tilde{K}_j^\dagger,
\label{eq:SingleStepWalk}
\end{eqnarray}
where $\tilde{K}_j = \hat{K}_j/\sqrt{\tilde{P}_j}$. This definition will easily allow us to define our protocol through the ensemble average of quantum trajectories.

We now suppose that we have a protocol (which in our case uses ancilla systems and quantum measurement), labeled $\tilde{\mathcal{B}}$, that implements one of the $\tilde{K}_j$ sampled from the set $\{\tilde{K}_j\}$ with the correct probability $\tilde{P}_j$. Then, if we implement this protocol many times on identical copies of the same initial state $\rho$, we have that
\begin{eqnarray}
        \mathbb{E}\left(\tilde{\mathcal{B}}\left[\rho\right]\right) = \sum_j \tilde{P}_j\tilde{K}_j\rho\tilde{K}_j^\dagger = \mathcal{B}[\rho],
\end{eqnarray}
where $\mathbb{E}\left(.\right)$ is the ensemble average. We will use the above description as a single time-step of the discrete-time QSW. 

Let us now consider repeated action of $\tilde{\mathcal{B}}$, and in analogy to the ``quantum trajectories on a quantum computer'' scheme developed in Ref.~\cite{Govia2017} we shall refer to each instance of such repeated action as a \emph{trajectory}. By linearity of the quantum operations, we see that
\begin{eqnarray}
        \mathbb{E}\left(\tilde{\mathcal{B}}\left[\tilde{\mathcal{B}}\left[\rho\right]\right]\right) = \mathbb{E}\left(\tilde{\mathcal{B}}\left[\mathbb{E}\left(\tilde{\mathcal{B}}\left[\rho\right]\right)\right]\right) = \mathcal{B}\left[\mathcal{B}[\rho]\right],
\end{eqnarray}
which can be trivially extended to any number of actions of $\tilde{\mathcal{B}}$. Thus, by averaging over the final outcome of many trajectories we can simulate the action of an arbitrary number of repetitions of the Kraus map $\mathcal{B}$. In the rest of this manuscript, we detail how to implement a map of the form of $\tilde{\mathcal{B}}$ for the case of a discrete-time QSW.

\section{Quantum Simulation of a Discrete-Time Quantum Stochastic Walk}
\label{sec:Protocol}
Let $\mathcal{G}=\left(V(\mathcal{G}),E(\mathcal{G})\right)$ be an arbitrarily connected (and possibly directed) graph with vertices $V(\mathcal{G})$ and edges $E(\mathcal{G})$, and $\{|n\rangle,1\leq n\leq |V(\mathcal{G}|\}$ a set of pairwise orthonormal quantum states which enumerate the location of a ``walker'' on the graph vertices. We will restrict our system to the single excitation subspace, so that $|n\rangle$ denotes a quantum state with a single excitation in vertex $n$ and all other vertices empty.

For any connected graph $\mathcal{G}$, we consider a quantum stochastic map $\mathcal{B}$ representing a single time-step of a QSW, which can be written in Kraus form as 
\begin{eqnarray}
\label{StochasticMap}
\mathcal{B}[\rho]:=\alpha\hat{U}_{\mathcal{G}}(\Delta t)\rho(t)\hat{U}_{\mathcal{G}}^{\dagger}(\Delta t)+\sum_{(m,n)\in E\left(\mathcal{G}\right)}\kappa_{nm}{|m\rangle\langle n|\rho|n\rangle\langle m|}.
\end{eqnarray}
Here, $\hat{U}_{\mathcal{G}}(\Delta t):=e^{-i\hat{H}_{\mathcal{G}}\Delta t}$ is the propagator of the graph coherent evolution for a time-step of length $\Delta t$, generated by the Hamiltonian $\hat{H}_{\mathcal{G}}$ of the graph $\mathcal{G}$. The coefficients $\alpha,\kappa_{nm}\in[0,1]$ represent the weights for coherent or incoherent processes to happen and satisfy $\sum_{m}\kappa_{nm}=1-\alpha$ for all $n\in V\left(\mathcal{G}\right)$  due to trace-preservation.

We define a discrete-time quantum stochastic walk by the repeated application of the single time-step quantum stochastic map $\mathcal{B}$ to the initial state $\rho_0$
\begin{eqnarray}
\label{DiscreteQSW}
\rho_n=\mathcal{B}^n\left[\rho_0\right]:=\underbrace{\mathcal{B}[\mathcal{B}[\ldots\mathcal{B}\left[\rho_0\right]}_{n\textnormal{ times}}\ldots]].
\end{eqnarray}
In the following, we show how to construct the quantum stochastic map $\tilde{\mathcal{B}}$ that, as described previously, can be used to simulate $\mathcal{B}$ via the ensemble average. To do this for any connected graph $\mathcal{G}$, we use its key building-block: the 2-vertex graph $\mathcal{G}_2$ with a single (possibly directed) edge. Notice that equation (\ref{StochasticMap}) is of the form of the single time-step quantum operation, see equation (\ref{eq:SingleStepWalk}). We thus need to define how to implement each of its Kraus operators.

\subsection{A general 2-vertex graph}
\label{sec:SingleEdge}

\begin{figure}
\includegraphics[width = 0.65\columnwidth]{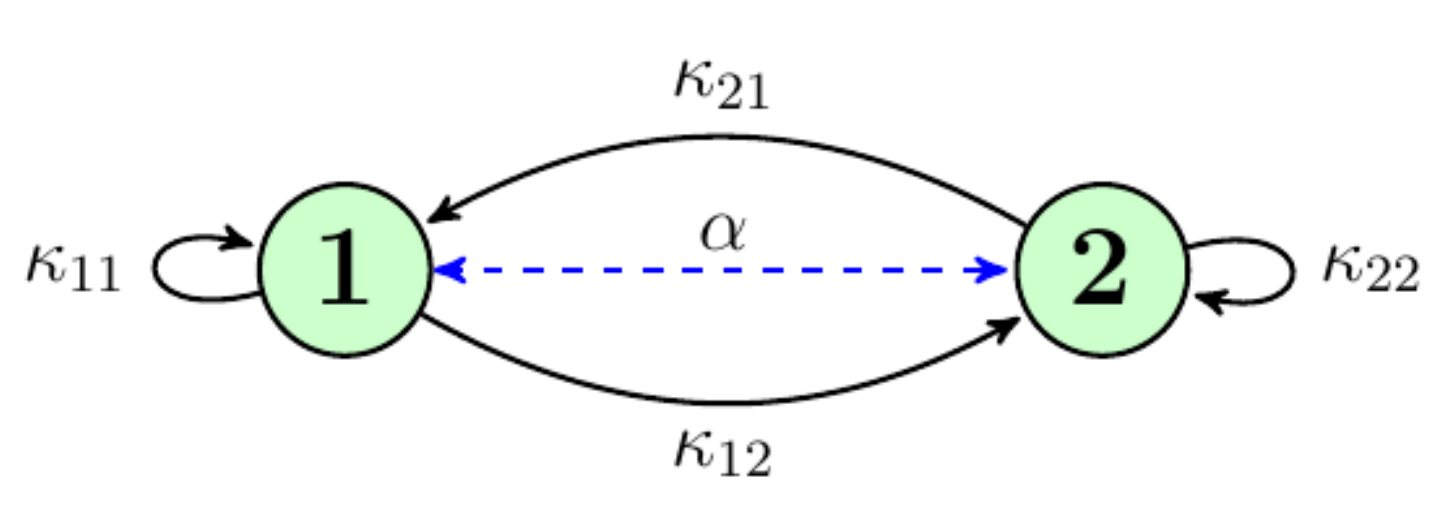}
\caption{The most general 2-vertex-graph as the key building-block for arbitrary graphs. The vertices are coherently coupled (blue dashed arrow) as well as incoherently coupled (black arrows). The probabilities satisfy $\kappa_{11}+\kappa_{12}=\kappa_{21}+\kappa_{22}=1-\alpha$ due to trace-preservation of the density matrix.}
\label{fig:Edge}
\end{figure}

Let us consider the most general 2-vertex graph $\mathcal{G}_2$ with coherent edge coupling and all possible directed\footnote{We call an edge directed if any of the $\kappa_{nm} > 0$, since that is the defining difference between a QSW and a coherent QW.} edges, see figure~\ref{fig:Edge}. As we will argue, the procedure outlined below easily generalizes to larger  graphs. For such 2-vertex graphs, equation (\ref{StochasticMap}) becomes
\begin{eqnarray}
\label{StochasticMap2}
\mathcal{B}[\rho]:=\alpha\hat{U}_{\mathcal{G}_2}(\Delta t)\rho(t)\hat{U}_{\mathcal{G}_2}^{\dagger}(\Delta t)+\sum_{m,n=1}^2\kappa_{nm}{|m\rangle\langle n|\rho|n\rangle\langle m|},
\end{eqnarray}
where trace-preservation implies that $\kappa_{11}+\kappa_{12}=\kappa_{21}+\kappa_{22}=1-\alpha$. The full system will be comprised of the original graph vertices, represented by the basis states $\ket n$, and one ancillary quantum state coupled to graph vertex. We shall refer to the graph vertices simply as the {\it system}, and the ancillary states as the {\it ancillae}. The ancillae will be used to implement the stochastic processes. A single time-step of the QSW, given by equation (\ref{StochasticMap2}), will be divided into three parts (see figure~\ref{fig:protocolSingleEdge}):
\begin{figure}
\begin{center}
\includegraphics[width = 0.65\columnwidth]{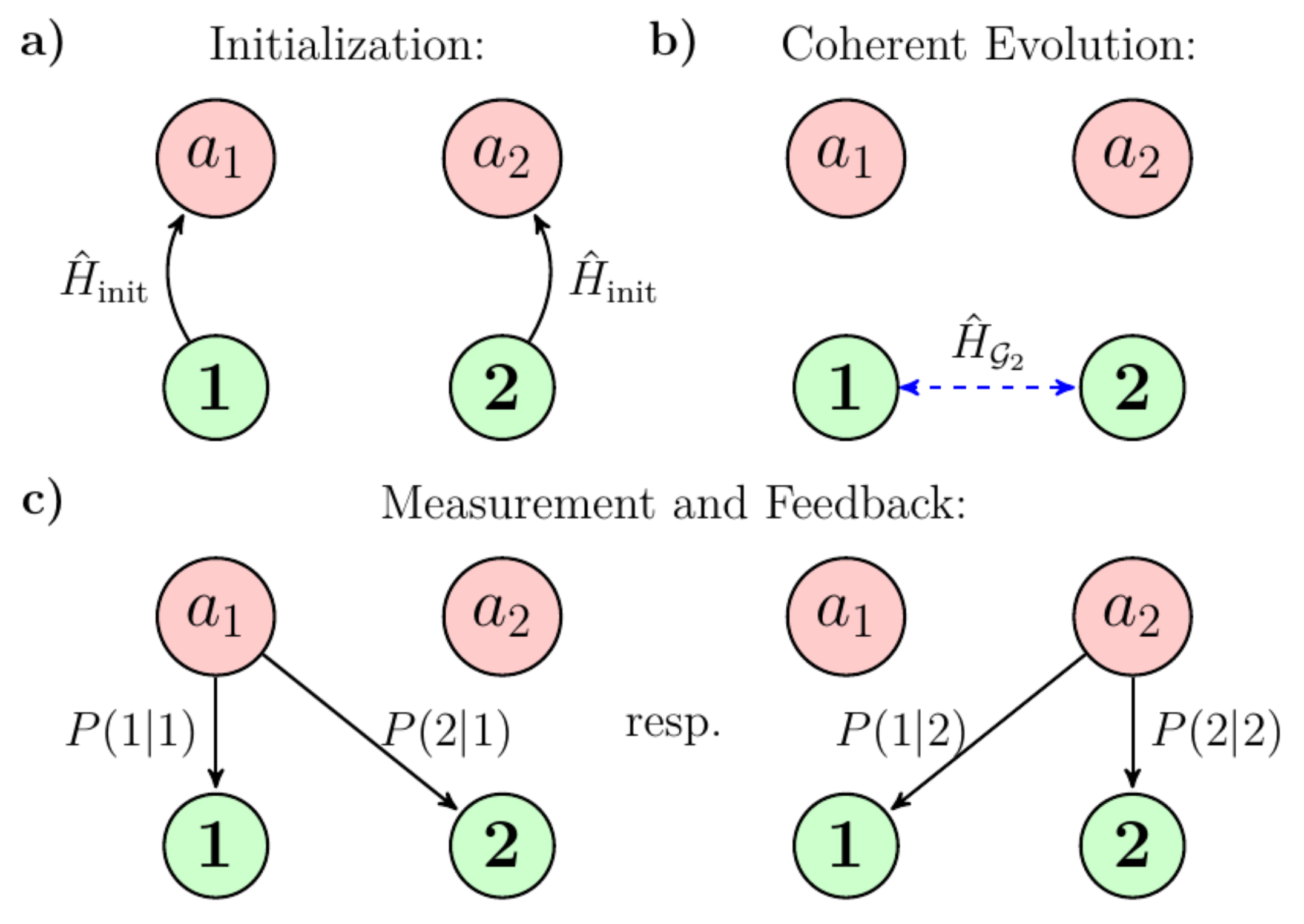}
\caption{Protocol to simulate a discrete-time quantum stochastic walk on the most general 2-vertex graph $\mathcal{G}_2$. \textbf{a)} The vertex-states are coherently coupled to their corresponding ancillae via $\hat{H}_{\rm init}$ for time $\Delta t_{\rm init}$ such that $g\Delta t_{\rm init}=\arccos\left(\sqrt{\alpha}\right)$. \textbf{b)} The vertex-states are coupled according to the coherent part of the graph $\mathcal{G}_2$. \textbf{c)}  The population of the ancillae is measured. If one of them is found to be occupied, the excitation is transitioned to one of the vertices according to the corresponding rates. If the ancillae are both found to be empty, no further feed-forward step is needed. This completes a single time-step of the discrete-time quantum stochastic walk.}
\label{fig:protocolSingleEdge}
\end{center}
\end{figure}

(1) {\it Initialization}: At the start of each time-step, the system and ancillae are uncoupled with no excitations in the ancillae. The density matrix can then be written as
\begin{eqnarray}
\label{eq:rho0}
\rho_0= \left(\begin{array}{rrrr} \rho_{11} & \rho_{12} & \rho_{1a_1} & \rho_{1a_2}\\
                \rho_{21} & \rho_{22} &  \rho_{2a_1} & \rho_{2a_2}\\
                \rho_{a_11} & \rho_{a_12} &  \rho_{a_1a_1} & \rho_{a_1a_2}\\
                \rho_{a_21} & \rho_{a_22} &  \rho_{a_2a_1} & \rho_{a_2a_2}\\
        \end{array}\right)= \left(\begin{array}{rrrr} \rho_{11} & \rho_{12} & 0 & 0\\
                \rho_{21} & \rho_{22} &  0 & 0\\
                0 & 0 &  0 & 0\\
                0 & 0 &  0 & 0\\
        \end{array}\right),
\end{eqnarray}
where the subscript 1 (2) denotes vertex 1 (2), and subscripts $a_1,~a_2$ denote the corresponding ancillae. All vertices are then coupled to their corresponding ancillae via the interaction Hamiltonian
\begin{eqnarray}
\hat{H}_{\rm init}:=g\left(|1\rangle\langle a_1|+|a_1\rangle\langle 1|+|2\rangle\langle a_2|+|a_2\rangle\langle 2|\right),
\end{eqnarray}
for time $\Delta t_{\rm init}$. All couplings are equal and $\Delta t_{\rm init}$ is chosen such that
\begin{eqnarray}
\label{Arccos}
g\Delta t_{\rm init}=\arccos\left(\sqrt{\alpha}\right).
\end{eqnarray}
This choice is crucial, as will become clear in the second step discussed below. It results in a density matrix $\rho_{\rm init}$ right after the initialization step that is given by
\begin{eqnarray}
\label{Initialization}
\rho_{\rm init}=\hat{U}_{\rm init}\rho_0\hat{U}^{\dagger}_{\rm init}=
\left(\begin{array}{cc}
\begin{array}{cc}
\alpha\rho_{11} & \alpha\rho_{12}\\
\alpha\rho_{21} & \alpha\rho_{22}
\end{array} & \Upsilon  \\
           \Upsilon & \Upsilon  \\
        \end{array}\right),
\end{eqnarray}
where $\hat{U}_{\rm init}=e^{-i\hat{H}_{\rm init}\Delta t_{\rm init}}$ and the $\Upsilon$ symbols represent generic $2\times2$ matrices whose precise form is not relevant at this stage. 

(2) {\it Coherent Evolution}: We now decouple the ancillae from the system and implement the desired coherent evolution between the graph vertices within the system
\begin{eqnarray}
\hat{H}_{\mathcal{G}_2}:=g_{\rm coh}\left(|1\rangle\langle 2|+|2\rangle\langle 1|\right)
\end{eqnarray}
for the desired length of the time-step $\Delta t$. Note that $\hat{H}_{\mathcal{G}_2}$ is the Hamiltonian of the graph $\mathcal{G}_2$, which has one free element $g_{\rm coh} \geq 0$. The density matrix $\rho_{\rm coh}$ after the coherent evolution is
\begin{eqnarray}
\rho_{\rm coh}=\hat{U}_{\mathcal{G}_2}(\Delta t)\rho_{\rm init}\hat{U}^{\dagger}_{\mathcal{G}_2}(\Delta t),
\end{eqnarray}
where $\hat{U}_{\mathcal{G}_2}=e^{-i\hat{H}_{\mathcal{G}_2}\Delta t}$ is the propagator of the graph Hamiltonian. Explicitly, at the end of the coherent evolution step we obtain

\begin{eqnarray}
\rho_{\rm coh}=
\left(\begin{array}{cc}
\alpha\hat{U}_{\mathcal{G}_2}(\Delta t)\left(\begin{array}{cc}
\rho_{11} & \rho_{12}\\
\rho_{21} & \rho_{22}
\end{array}\right)\hat{U}^{\dagger}_{\mathcal{G}_2}(\Delta t) & \Upsilon  \\
           \Upsilon & \Upsilon  \\
        \end{array}\right).
\end{eqnarray}
which is an implementation of the first term (coherent evolution) on the right-hand side of equation (\ref{StochasticMap2}) in the subspace of the vertices.

(3) {\it Measurement and Feed-Forward}:
The final part of the protocol uses quantum measurement to randomly determine which term from equation (\ref{StochasticMap2}) is implemented for each time-step in a given trajectory. The coherence between vertices and ancillae is also removed, guaranteeing that once each time-step is concluded the system state is of the form of equation (\ref{eq:rho0}). To do this, we decouple the system vertices and measure all ancillae simultaneously. As the system is restricted to the single-excitation subspace, there are three possible results:
\begin{enumerate}
\item the excitation is measured in ancilla $a_1$,
\item the excitation is measured in ancilla $a_2$,
\item all ancillae are found empty.
\end{enumerate}
For an $n$-vertex graph there are $n+1$ possible measurement outcomes. The last outcome guarantees that the walker is in one of the system vertices, and the second step implements the coherent evolution part of equation (\ref{StochasticMap2}) in this case. No further action is required, and we can proceed to the next time-step in the trajectory.

The other measurement results are interpreted as one of the incoherent processes having taken place. To determine which, we use the incoherent rates $\kappa_{ij}$ of the intended QSW as follows. If the excitation is found in ancilla $a_i$ this fixes the index $i$ in $\kappa_{ij}$, i.e.~the starting vertex of the incoherent process. To determine the index $j$ and implement the incoherent evolution, we randomly choose $j$ from a probability distribution given by the conditional probabilities $P(j|i)$, and then move the excitation to system vertex $j$. These conditional probabilities are given by
\begin{eqnarray}
\label{CondProbs}
P(j|i)=\frac{\kappa_{ij}}{\sum_j\kappa_{ij}}.
\end{eqnarray}
In this feed-forward operation, the outcome of the quantum measurement combined with the outcome of the classical random choice determines which of the incoherent terms in equation (\ref{StochasticMap2}) is implemented in this time-step of the trajectory.

The complete set of operators which describe the measurement and feed-forward step for a two-vertex graph is given by $\big\{\hat{M}_0,\hat{M}^{a_1}_{1},\hat{M}^{a_1}_{2},\hat{M}^{a_2}_{1},\hat{M}^{a_2}_{2}\big\}$, with matrix representations given in~\ref{app:Matrices}. $\hat{M}_0$ describes the measurement outcome where both ancillae are found to be empty, and we write
\begin{eqnarray}
\hat{M}^{a_1}_{1/2}=\hat{F}^{a_1}_{1/2}\hat{M}_{a_1}\textnormal{ and }\hat{M}^{a_2}_{1/2}=\hat{F}^{a_2}_{1/2}\hat{M}_{a_2},
\end{eqnarray}
where $\hat{M}_{a_1/a_2}$ describes the measurement where the excitation is found in ancilla $a_1/a_2$, and $\hat{F}^{a_1/a_2}_{1/2}$ describes the conditional feed-forward according to the measurement result and classical random choice.

The three step procedure outlined above implements a single step of a single trajectory of the discrete-time QSW. Averaging over many trajectories we obtain the density matrix
\newpage
\begin{eqnarray}
\label{Rhof4x4}
\rho_{\Delta t}&:=\hat{M}_0\rho_{\rm coh}\hat{M}^{\dagger}_0+\sum_{y\in\{1,2\}}\sum_{x\in\{a_1,a_2\}}\hat{M}^{x}_{y}\rho_{\rm coh}{\hat{M}^{x\dagger}_{y}}\\
&=\left(\begin{array}{rr}
\mathcal{B}\left[\left(\begin{array}{rr} \rho_{11} & \rho_{12}\\ \rho_{21} & \rho_{22}\end{array}\right)\right] &\begin{array}{rr} 0 & 0\\ 0 & 0\end{array}\\ \begin{array}{rr} 0 & 0\\ 0 & 0\end{array}\hspace{.8cm} & \begin{array}{rr} 0 & 0\\ 0 & 0\end{array}\\
\end{array}\right).
\end{eqnarray}
A $k$-step trajectory is performed by $k$ successive implementations of the above protocol, with its ensemble average having the desired statistics to simulate the QSW.

\subsection{Arbitrary Graphs}
\label{sec:ArbitraryGraphs}
The protocol proposed in section \ref{sec:SingleEdge} generalizes trivially to any larger graph $\mathcal{G}$, with each system vertex requiring an ancilla. As before, a single time-step is split into three parts:

(1) {\it Initialization}: System states are coupled to their corresponding ancillae via
\begin{eqnarray}
\label{Hinit}
\hat{H}_{\rm init}:=\sum_{m\in V\left(\mathcal{G}\right)}g\left(|m\rangle\langle a_m|+|a_m\rangle\langle m|\right),
\end{eqnarray}
for a time $\Delta t_{\rm init}$. Here, the summation covers all the graph vertices $m\in V\left(\mathcal{G}\right)$ and the state $|a_m\rangle$ denotes the ancilla state which corresponds to vertex $m$. Again, $\Delta t_{\rm init}$ is chosen such that
\begin{eqnarray}
g\Delta t_{\rm init}=\arccos\left(\sqrt{\alpha}\right).
\end{eqnarray}

(2) {\it Coherent Evolution}: The ancillae are now decoupled from the system and the system evolves coherently with
\begin{eqnarray}
\label{JCH}
\hat{H}_{\mathcal{G}}:=\sum_{(n,m)\in E\left(\mathcal{G}\right)}g_{nm}\left(|m\rangle\langle n|+|n\rangle\langle m|\right),
\end{eqnarray}
for a time $\Delta t$. Note that equation (\ref{JCH}) is the full Hamiltonian of the graph as the summation covers all the edges of the graph.

(3) {\it Measurement and Feed-Forward}: Finally, the ancillae are measured. As before, if the ancilla are all found to be empty the time-step is complete. If the excitation is found in an ancilla, then the excitation will be incoherently moved to a randomly chosen system vertex that is connected to the system vertex corresponding to the excited ancilla. This process is identical to that described previously for a two-vertex graph, but with the choice of final vertex expanded to include all vertices connected incoherently ($\kappa_{ij} > 0$) to the initial vertex.

Parts (1)-(3) implement a single time-step of equation (\ref{StochasticMap2}). Again, the complete walk will be given by iterating this procedure. This simple generalization is possible for two main reasons: (i) each graph vertex is only coupled to a single ancilla, and (ii) the ancillae are never directly coupled to each other.

\section{Simulating a Continuous-Time QSW by a Discrete-Time QSW}
\label{sec:ApproximateTimeEvolution}

In the previous sections we showed how to simulate discrete-time quantum stochastic walks as defined by equations (\ref{StochasticMap}) and (\ref{DiscreteQSW}), using a trajectory approach. However, currently the majority of applications of QSWs use the continuous-time version, as is widely documented in literature~\cite{Schuld2014c,Cuevas,Mohseni2008,Zimboras:2013aa,Caruso:2014aa,Viciani:2015aa,Caruso:2016aa,Park:2016aa}. Therefore, we now show how to implement a restricted set of continuous-time QSW by a discrete-time QSW, such that our method for simulating discrete-time QSWs is also applicable.

The continuous-time quantum stochastic walk of Ref.~\cite{Whitfield2010} is given by a Lindblad master equation of the form
\begin{eqnarray}
\label{QSW}
\dot{\rho}= (\omega-1)i\left[\hat{H}_{\mathcal{G}},\rho\right] + \omega\sum_{k}\gamma_k{\left(\hat{L}_k\rho\hat{L}_k^{\dagger}-\frac{1}{2}\big\{\hat{L}_k^{\dagger}\hat{L}_k,\rho\big\}\right)},
\end{eqnarray}
where $\rho$ is the density operator of the system, $\hat{L}_k$ are the Lindblad operators with $\gamma_k$ their associated incoherent transition rates, $\hat{H}_{\mathcal{G}}$ is the Hamiltonian of the underlying graph $\mathcal{G}$ and $\omega\in[0,1]$.  For $\omega=0$, we obtain the completely coherent quantum walk and for $\omega=1$, the classical random walk. Hence, for $\omega\in(0,1)$, equation (\ref{QSW}) leads to dynamics we could not obtain in a purely coherent or incoherent framework. 

We write the Liouvillian $\mathcal{L}_{\omega}$ of equation (\ref{QSW}) as 
\begin{eqnarray}
\mathcal{L}_{\omega}\rho=(1-\omega)\mathcal{H}\rho+\omega\Lambda\rho
,\label{Liouvillian}
\end{eqnarray}
where $\mathcal{H}\rho=-i\left[\hat{H}_{\mathcal{G}},\rho\right]$ and
\begin{eqnarray}
\Lambda\rho=\sum_{k}\gamma_k{\left(\hat{L}_k\rho\hat{L}_k^{\dagger}-\frac{1}{2}\big\{\hat{L}_k^{\dagger}\hat{L}_k,\rho\big\}\right)}.
\end{eqnarray}
The Liouvillian (\ref{Liouvillian}) is the generator of a quantum dynamical semigroup~\cite{Breuer:2006uq}. Therefore, we can write
\begin{eqnarray}
\rho(t+\Delta t)=\exp\left(\mathcal{L}_{\omega}\Delta t\right)\rho(t)=\left(\sum_{l=0}^{\infty}\frac{1}{l!}\mathcal{L}_{\omega}^l\Delta t^l\right)\rho(t).
\label{Semigroup}
\end{eqnarray}
Inserting equation (\ref{Liouvillian}) into equation (\ref{Semigroup}) yields, up to first order in $\Delta t$,
\begin{eqnarray}
\fl\qquad\qquad\rho(t+\Delta t)&=\Big(1+\Delta t\left((1-\omega)\mathcal{H}_{\mathcal{G}}+\omega\Lambda\right)\nonumber +\mathcal{O}\left(\Delta t^2\right)\Big)\rho(t)\nonumber \\&=\Big((1-\omega)\left(1+\Delta t\mathcal{H}_{\mathcal{G}}\right)+\omega\left(1+\Delta t\Lambda\right)+\mathcal{O}\left(\Delta t^2\right)\Big)\rho(t).
\label{FirstOrderApproximation}
\end{eqnarray}
The first term on the right-hand side of equation (\ref{FirstOrderApproximation}) can be interpreted as pure coherent evolution that occurs with probability $(1-\omega)$. Analogously, we interpret the second term as describing incoherent evolution occurring with probability $\omega$.

We now consider continuous-time QSWs where the incoherent evolution describes incoherent excitation transfer between system vertices~\cite{Cuevas}, such that $\hat{L}_k = \left|m\rangle\langle n\right|$. In this case, the incoherent evolution of equation (\ref{FirstOrderApproximation}) becomes
\begin{eqnarray}
\fl\qquad\omega\left(1+\Delta t\Lambda\right)\rho 
=\omega\rho+\sum_{(m,n)\in E\left(\mathcal{G}\right)}\omega\Delta t\gamma_{nm}{\left(|m\rangle\langle n|\rho|n\rangle\langle m|-\frac{1}{2}\big\{|n\rangle\langle n|,\rho\big\}\right)}.
\end{eqnarray}
Here, the $k$-th Lindblad operator $\hat{L}_k=|m\rangle\langle n|$ generates an incoherent jump from vertex $n$ to vertex $m$, and $\gamma_{nm}$ describes the transition rate for this process, with $E\left(\mathcal{G}\right)$ the set of connected edges of the graph.

To first order in $\Delta t$, we see that $p_{nm} =\Delta t\gamma_{nm}$ can be treated as the conditional probability to transition to vertex $m$ if the excitation is in vertex $n$ during time-step $\Delta t$, if the Lindblad rates satisfy
\begin{eqnarray}
\label{CondProb}
\sum_{m\in V\left(\mathcal{G}\right)}p_{nm}= \sum_{m\in V\left(\mathcal{G}\right)}\Delta t\gamma_{nm} =1, \label{eqn:probcon}
\end{eqnarray}
to ensure conservation of probability. This condition must be satisfied simultaneously for all $n$, which is possible if and only if
\begin{eqnarray}
\sum_{m\in V\left(\mathcal{G}\right)}\gamma_{nm}=\gamma\hspace{1cm}\forall n\in V\left(\mathcal{G}\right),
\end{eqnarray}
such that $\Delta t=\gamma^{-1}$ can be chosen uniquely to simultaneously guarantee equation (\ref{eqn:probcon}) for all $n$. We note that this is the same restriction on the Lindblad rates as was necessary for protocols to simulate continuous time QSWs using quantum trajectories on a quantum computer \cite{Govia2017}.

Under this restriction, we can write the incoherent evolution as
\begin{eqnarray}
\nonumber\fl\omega\rho+\omega\sum_{(m,n)\in E\left(\mathcal{G}\right)}p_{nm} {\left(|m\rangle\langle n|\rho|n\rangle\langle m|-\frac{1}{2}\big\{|n\rangle\langle n|,\rho\big\}\right)}
\\ \fl\nonumber=\omega\rho+\omega\sum_{(m,n)\in E\left(\mathcal{G}\right)}p_{nm} |m\rangle\langle n|\rho|n\rangle\langle m| -\frac{\omega}{2} \underbrace{\sum_{n\in V\left(\mathcal{G}\right)}\big\{|n\rangle\langle n|,\rho\big\}}_{=2\rho}\underbrace{\sum_{m\in V\left(\mathcal{G}\right)}p_{nm} }_{=1}
\\ \fl \label{eqn:kappadef} =\sum_{(m,n)\in E\left(\mathcal{G}\right)}\kappa_{nm}|m\rangle\langle n|\rho|n\rangle\langle m|,
\end{eqnarray}
where in the last line we have defined $\kappa_{nm} = \omega p_{nm}$. Thus, we see that the short-time incoherent evolution for this restricted class of continuous-time QSWs has the same form of the incoherent part of the discrete-time QSW we have used throughout this manuscript.

Similarly, we replace $\left(1+\Delta t\mathcal{H}_{\mathcal{G}}\right)$ with the unitary propagator $\hat{U}_{\mathcal{G}}(\Delta t) =e^{-i\hat{H}_{\mathcal{G}}\Delta t} $ for the coherent evolution. Combining these results, and defining $\alpha=1-\omega$, we see that we can write the continuous-time evolution for short $\Delta t$ as
\begin{eqnarray}
\label{ApproximationFinal}
\fl\rho(t+\Delta t)=\alpha\hat{U}_{\mathcal{G}}(\Delta t)\rho(t)\hat{U}_{\mathcal{G}}^{\dagger}(\Delta t) + \sum_{(m,n)\in E\left(\mathcal{G}\right)}\kappa_{nm}{|m\rangle\langle n|\rho|n\rangle\langle m|}+\mathcal{O}\left(\Delta t^2\right).
\end{eqnarray}
This has the form of a Kraus map for a discrete-time QSW, and as such, we have shown how to implement the short time evolution of a restricted class of continuous-time QSWs with a discrete-time QSW, broadening the applicability of our simulation method for discrete-time QSWs.

\section{Conclusion}
\label{sec:conc}
In this work, we developed a trajectory-based protocol to simulate discrete-time QSWs on a coherent quantum computer. This ancilla-based protocol breaks down each time-step of the QSW into three parts that require only coherent couplings, measurements, and feed-forward operations, and thus are suitable to implementation on quantum hardware. Subsequent applications of this process create a single quantum trajectory, and we show that, as with the standard quantum trajectories approach, ensemble averages over many trajectories mimics the desired QSW dynamics.

The full time-step was carefully detailed for the most general graph of two vertices and we have shown that this serves as a building block to simulate arbitrary graphs. The simple procedure to generalize to complex graphs is one of the key features of our proposal, as no complicated design of system-ancillae interaction is needed. The protocol can also be employed for simulations of continuous-time QSWs satisfying certain conditions, which are also present in previously proposed simulation methods using quantum computers.

We note that as our protocol is designed on the single-excitation subspace, hardware implementations using qubits to represent vertices and ancillae are not resource efficient. Such qubit implementations use a $2^N$-dimensional Hilbert space to simulate a graph $\mathcal{G}$ with $|V\left(\mathcal{G}\right)|=N$, which only requires $2N$ degrees of freedom including ancillae. An alternative approach could use qutrits to represent each vertex, with the third energy level representing the ancillae.

As is usual for quantum trajectory based protocols, our proposal will be more suited for simulations for which convergence scales faster than $d^2$, where $d$ is the Hilbert space dimension of the system to be simulated. As this heavily depends on the underlying graph $\mathcal{G}$, no general statement is possible. However, we note that as the system evolution is reset to a specific state after any ancilla is measured to be occupied, the protocol requires coherence times much shorter than the total simulation time and could therefore be useful for near-term quantum hardware implementations.

\ack
B.G.T. acknowledges support from FAPESC and CNPq INCT-IQ (465469/2014- 0).

\appendix

\section{Matrix representations of the measurement operators for a single edge}
\label{app:Matrices}
The matrix representations of the measurement operators in Section \ref{sec:SingleEdge} for a 2-vertex graph are:
\begin{eqnarray}
\fl\hat{M}_0=\left(\begin{array}{rrrr} 1 & 0 & 0 & 0\\
               0 & 1 & 0 & 0 \\
               0 & 0 & 0 & 0 \\
               0 & 0 & 0 & 0 \\
        \end{array}\right)\hspace{.8cm}\hat{M}_{a_1/a_2}=\left(\begin{array}{rrrr} 0 & 0 & 0 & 0\\
               0 & 0 & 0 & 0 \\
               0 & 0 & 1/0 & 0 \\
               0 & 0 & 0 & 0/1 \\
        \end{array}\right)\\
\fl\hat{F}^{a_1}_{1}=P(1|1)\left(\begin{array}{rrrr} 0 & 0 & 1 & 0\\
               0 & 1 & 0 & 0 \\
               1 & 0 & 0 & 0 \\
               0 & 0 & 0 & 1 \\
        \end{array}\right)\hspace{.8cm}\hat{F}^{a_1}_{2}=P(2|1)\left(\begin{array}{rrrr} 1 & 0 & 0 & 0\\
               0 & 0 & 1 & 0 \\
               0 & 1 & 0 & 0 \\
               0 & 0 & 0 & 1 \\
        \end{array}\right)\\
\fl\hat{F}^{a_1}_{1}=P(1|2)\left(\begin{array}{rrrr} 0 & 0 & 0 & 1\\
               0 & 1 & 0 & 0 \\
               0 & 0 & 1 & 0 \\
               1 & 0 & 0 & 0 \\
        \end{array}\right)\hspace{.8cm}\hat{F}^{a_2}_{2}=P(2|2)\left(\begin{array}{rrrr} 1 & 0 & 0 & 0\\
               0 & 0 & 0 & 1 \\
               0 & 0 & 1 & 0 \\
               0 & 1 & 0 & 0 \\
        \end{array}\right)
\end{eqnarray}

\section*{References}
\bibliographystyle{unsrt}

\bibliography{BibQuantumWalks}

\begin{thebibliography}{10}

\bibitem{Aharonov1993}
Y.~Aharonov, L.~Davidovich, and N.~Zagury.
\newblock Quantum random walks.
\newblock {\em Phys. Rev. A}, 48:1687--1690, Aug 1993.

\bibitem{Farhi1998}
Edward Farhi and Sam Gutmann.
\newblock {Quantum computation and decision trees}.
\newblock {\em Physical Review A}, 58(2):915--928, August 1998.

\bibitem{Aharonov2001}
Dorit Aharonov, Andris Ambainis, Julia Kempe, and Umesh Vazirani.
\newblock {Quantum walks on graphs}.
\newblock In {\em Proceedings of the thirty-third annual ACM symposium on
  Theory of computing - STOC '01}, pages 50--59, New York, New York, USA, July
  2001. ACM Press.

\bibitem{ambainis2001one}
Andris Ambainis, Eric Bach, Ashwin Nayak, Ashvin Vishwanath, and John Watrous.
\newblock One-dimensional quantum walks.
\newblock In {\em Proceedings of the thirty-third annual ACM symposium on
  Theory of computing}, pages 37--49. ACM, 2001.

\bibitem{Childs2003}
Andrew~M. Childs, Richard Cleve, Enrico Deotto, Edward Farhi, Sam Gutmann, and
  Daniel~A. Spielman.
\newblock {Exponential algorithmic speedup by a quantum walk}.
\newblock In {\em Proceedings of the thirty-fifth ACM symposium on Theory of
  computing - STOC '03}, page~59, New York, New York, USA, June 2003. ACM
  Press.

\bibitem{Lovett2010}
Neil~B. Lovett, Sally Cooper, Matthew Everitt, Matthew Trevers, and Viv Kendon.
\newblock Universal quantum computation using the discrete-time quantum walk.
\newblock {\em Phys. Rev. A}, 81:042330, Apr 2010.

\bibitem{Schuld2014c}
Maria Schuld, Ilya Sinayskiy, and Francesco Petruccione.
\newblock {Quantum walks on graphs representing the firing patterns of a
  quantum neural network}.
\newblock {\em Physical Review A}, 89(3):032333, mar 2014.

\bibitem{Rebentrost2014}
Patrick Rebentrost, Masoud Mohseni, and Seth Lloyd.
\newblock Quantum support vector machine for big data classification.
\newblock {\em Phys. Rev. Lett.}, 113:130503, Sep 2014.

\bibitem{Shenvi2003}
Neil Shenvi, Julia Kempe, and K.~Birgitta Whaley.
\newblock Quantum random-walk search algorithm.
\newblock {\em Physical Review A}, 67(5), may 2003.

\bibitem{Mohseni2008}
Masoud Mohseni, Patrick Rebentrost, Seth Lloyd, and Al{\'{a}}n Aspuru-Guzik.
\newblock {Environment-assisted quantum walks in photosynthetic energy
  transfer.}
\newblock {\em The Journal of chemical physics}, 129(17):174106, nov 2008.

\bibitem{Walschaers:2013aa}
Mattia Walschaers, Jorge Fernandez-de-Cossio Diaz, Roberto Mulet, and Andreas
  Buchleitner.
\newblock Optimally designed quantum transport across disordered networks.
\newblock {\em Phys. Rev. Lett.}, 111:180601, Oct 2013.

\bibitem{Whitfield2010}
James~D. Whitfield, C\'esar~A. Rodr\'{i}guez-Rosario, and Al\'an Aspuru-Guzik.
\newblock Quantum stochastic walks: A generalization of classical random walks
  and quantum walks.
\newblock {\em Phys. Rev. A}, 81:022323, Feb 2010.

\bibitem{Taketani2016}
Bruno~G. Taketani, Luke C.~G. Govia, and Frank~K. Wilhelm.
\newblock Physical realizability of continuous-time quantum stochastic walks.
\newblock {\em Phys. Rev. A}, 97:052132, May 2018.

\bibitem{Govia2017}
Luke C~G Govia, Bruno~G Taketani, Peter~K Schuhmacher, and Frank~K Wilhelm.
\newblock Quantum simulation of a quantum stochastic walk.
\newblock {\em Quantum Science and Technology}, 2(1):015002, jan 2017.

\bibitem{CrispinGardiner2004}
Crispin Gardiner and Peter Zoller.
\newblock {\em Quantum Noise}.
\newblock Springer Berlin Heidelberg, 2004.

\bibitem{Nielsen2000}
M.A. Nielsen and I.L. Chuang.
\newblock {\em Quantum Computation and Quantum Information}.
\newblock Cambridge University Press, Cambridge, UK, 2000.

\bibitem{Cuevas}
Hans~J Briegel and Gemma De~las Cuevas.
\newblock Projective simulation for artificial intelligence.
\newblock {\em Scientific reports}, 2, 2012.

\bibitem{Zimboras:2013aa}
Zolt{\'a}n Zimbor{\'a}s, Mauro Faccin, Zolt{\'a}n K{\'a}d{\'a}r, James~D.
  Whitfield, Ben~P. Lanyon, and Jacob Biamonte.
\newblock Quantum transport enhancement by time-reversal symmetry breaking.
\newblock {\em Scientific Reports}, 3:2361 EP --, 08 2013.

\bibitem{Caruso:2014aa}
Filippo Caruso.
\newblock Universally optimal noisy quantum walks on complex networks.
\newblock {\em New Journal of Physics}, 16(5):055015, 2014.

\bibitem{Viciani:2015aa}
Silvia Viciani, Manuela Lima, Marco Bellini, and Filippo Caruso.
\newblock Observation of noise-assisted transport in an all-optical
  cavity-based network.
\newblock {\em Phys. Rev. Lett.}, 115:083601, Aug 2015.

\bibitem{Caruso:2016aa}
Filippo Caruso, Andrea Crespi, Anna~Gabriella Ciriolo, Fabio Sciarrino, and
  Roberto Osellame.
\newblock Fast escape of a quantum walker from an integrated photonic maze.
\newblock {\em Nat Commun}, 7, 06 2016.

\bibitem{Park:2016aa}
Heechul Park, Nimrod Heldman, Patrick Rebentrost, Luigi Abbondanza, Alessandro
  Iagatti, Andrea Alessi, Barbara Patrizi, Mario Salvalaggio, Laura Bussotti,
  Masoud Mohseni, Filippo Caruso, Hannah~C. Johnsen, Roberto Fusco, Paolo
  Foggi, Petra~F. Scudo, Seth Lloyd, and Angela~M. Belcher.
\newblock Enhanced energy transport in genetically engineered excitonic
  networks.
\newblock {\em Nat Mater}, 15(2):211--216, 02 2016.

\bibitem{Breuer:2006uq}
Heinz-Peter Breuer and Francesco Petruccione.
\newblock {\em The Theory of Open Quantum Systems}.
\newblock Oxford University Press, 2006.

\end{thebibliography}

\end{document}